\documentclass[onecolumn,,showpacs,preprintnumbers,amsmath,amssymb,superscriptaddress,nofootinbib, preprint,12pt]{revtex4}
\usepackage{epsfig}
\usepackage{CJK}
\usepackage{amsfonts}
\usepackage{graphicx}
\usepackage{epsfig}
\usepackage{eepic}
\usepackage{amsmath}
\usepackage{amssymb}
\usepackage{color}
\usepackage{bbm}
\usepackage{dcolumn}
\usepackage{bm}
\usepackage{ulem}
\usepackage{slashbox}

\def\bc{\begin{center}}
\def\nno{\nonumber}
\def\ec{\end{center}}
\def\be{\begin{eqnarray}}
\def\ee{\end{eqnarray}}
\definecolor{dyellow}{rgb}{1.,0.8,.0}
\definecolor{myblue}{rgb}{.1,.1,.7}
\definecolor{dcyan}{rgb}{.0,.6,.6}
\definecolor{dmagenta}{rgb}{0.6,0.0,0.6}
\definecolor{brown}{rgb}{0.6,0.2,0.}
\definecolor{darkblue}{rgb}{.0,.0,0.5}
\definecolor{darkred}{rgb}{0.75,0.0,0.0}
\definecolor{orange}{rgb}{1.,.6,.0}
\definecolor{dorange}{rgb}{0.8,.4,.0}
\definecolor{darkgreen}{rgb}{0.0,0.6,0.0}
\definecolor{purple}{rgb}{.4,.0,.4}
\definecolor{lightgrey}{rgb}{0.7, 0.7, 0.7}
\definecolor{grey}{rgb}{0.4, 0.4, 0.4}

\def\blue{\color{blue}}

\def\al{\alpha}

\def\eps{\epsilon}

\def\la{\lambda}

\newcommand{\nc}{\newcommand}
\nc{\rnc}{\renewcommand} \nc{\ket}[1]{\left | \, #1 \right \rangle}
\nc{\bra}[1]{\left \langle #1 \, \right |}
\nc{\ua}{\uparrow} \nc{\da}{\downarrow}

\nc{\braket}[2]{\langle\, #1\,|\,#2\,\rangle}
\nc{\half}{\frac{1}{2}}

\nc{\prj}{\mathcal{P}} \nc{\hilb}{\mathcal{H}}
\nc{\pth}{\mathcal{C}} \nc{\inprod}[2]{\braket{#1}{#2}}
\nc{\upket}{\ket{\uparrow}} \nc{\downket}{\ket{\downarrow}}
\nc{\upbra}{\bra{\uparrow}} \nc{\downbra}{\bra{\downarrow}}

\begin{document}
\begin{CJK*}{GBK}{song}

\title{Further studies on holographic insulator/superconductor phase transitions from Sturm-Liouville eigenvalue problems  }
\author{Huai-Fan Li} \email{huaifan.li@stu.xjtu.edu.cn}
\affiliation{Institute of Theoretical Physics, Department of Physics, Shanxi Datong University, Datong, 037009, China}

 \begin{abstract}
 We take advantage of the Sturm-Liouville eigenvalue problem to analytically study the holographic insulator/superconductor phase transition in the probe limit. The interesting point is that this analytical method can not only estimate the most stable mode of the phase transition, but also the second stable mode. We find that this analytical method perfectly matches with other numerical methods, such as the shooting method. Besides, we argue that only Dirichlet boundary condition of the trial function is enough under certain circumstances, which will lead to a more precise estimation. This relaxation for the boundary condition of the trial function is first observed in this paper as far as we know.
 \end{abstract}


 \maketitle

\section{Introduction}
 The AdS/CFT correspondence \cite{Maldacena:1997re,Gubser:1998bc,Witten:1998qj} states that the gravity in an asymptotically AdS spacetime is dual to a field theory which sits on the boundary of the gravity. Although originally AdS/CFT correspondence was realized between a type IIB  supergravity with geometry $AdS^5\times S^5$ and its dual field theory with $\mathcal{N}=4$ super-Yang Mills on its boundary, the AdS/CFT correspondence has been widely extended to various general cases. For example, one can have a black hole in the bulk which is corresponding to a field theory with temperature on the boundary \cite{Witten:1998zw}; or one can work in an asymptotic Lifshitz spacetime \cite{Son:2008ye} which is dual to a non-relativistic field theory, etc. In recent years, the applications of AdS/CFT had an overwhelming development in condensed matter physics. For instance, the holographic superconductor(superfluids) phase transition \cite{Hartnoll:2008vx}, the holographic (non-)Fermi liquid \cite{Lee:2008xf,hongliu,zaanen}, and holographic insulator/superconductor phase transition \cite{Nishioka:2009zj,montull:2011pr,Roychowdhury:2013dr,Montull:2012pr}, etc. Among these, the holographic insulator/superconductor phase transition was studied in an AdS soliton spacetime \cite{Horowitz:1998ha} which is dual to a confined gauge field theory with a mass gap on the boundary \cite{Witten:1998zw}. This mass gap resembles the mass gap in the insulator in condensed matter physics, which is the motivation to use AdS soliton as the background. The insulator/superconductor phase transition occurs at some critical chemical potential, beyond which the charged scalar will have a non-vanishing value which spontaneously breaks the U(1) symmetry and then induces the superconducting phenomenon.

  A lot of numerical methods are developed to study the applications of the AdS/CFT correspondence, because in this case one needs to solve the coupled equations of motions of the system, which is difficult to study by analytical methods. In particular, the shooting method \cite{Hartnoll:2008vx} is adopted to study the holographic superconductor phase transition, in which one requires the source term of the scalar field be vanishing on the boundary. However, this does not mean that the analytical methods have no roles in the holographic studies. In \cite{Siopsis:2010uq}, the authors used the Sturm-Liouville eigenvalue problem (SL method for short) to analytically study the holographic superconductor phase transition. After this, a large number of papers, for example \cite{Zeng:2010zn,Li:2011xja,Cai:2011ky,Cai:2011tm,Cai:2011qm,Momeni:2011mo,Momeni:2012mo}, appeared to analytically study the holographic systems by virtue of the SL method. In particular, the authors in \cite{Cai:2011qm}  have investigated the holographic insulator/superconductor phase transition by three different methods, {\it i.e.}, the analytical SL method, the numerical shooting method and the quasinormal mode(QNM) methods. They found that the final results obtained from these three methods were in good agreement with each other for finding the most stable modes (vacuum solutions) of the phase transition. They also studied the second stable mode and the third stable mode by virtue of the last two methods, {\it viz},  the numerical shooting method and the QNM methods. But they did not use the first method, {\it i.e.}, the analytical SL method, to study the second mode of the phase transition which is the main purpose of this paper.

In this paper, we will extend our study on the holographic insulator/superconductor phase transition by virtue of the analytical SL method. Explicitly, we study the holographic insulator/superconductor phase transition for s-wave and p-wave in the probe limit in the AdS soliton background.  In particular, we find that this analytical SL method can not only estimate the most stable mode of the phase transition like in \cite{Cai:2011ky,Cai:2011qm}, but also can be applied to estimate the second stable mode. The critical chemical potentials for the second stable mode we obtained from the analytical SL method are in good agreement with those obtained from the shooting method and QNM method in \cite{Cai:2011qm}. In addition, we also argue that actually one can relax the boundary conditions of the trial function $F(z)$ to be only of Dirichlet in some circumstances, rather than imposing both Dirichlet and Nuemann boundary conditions as in the previous papers \cite{Siopsis:2010uq,Cai:2011ky}. We notice that this kind of relaxation of the boundary condition for the trial function will make the estimation for the critical chemical potential more precise than before, which is first observed in this paper as far as we know. Furthermore, we argue that this analytical SL  method with the one parameter trial function can hardly be used to estimate the third stable mode of the phase transition, because of the conditions of the extremal values for a smooth function. Therefore, this is the limitation of the analytical SL method compared to the shooting method or QNM method.

This paper is organized as follows: We will briefly review the Sturm-Liouville eigenvalue problem in Sec.\ref{sec:sl}; In Sec.\ref{sec:swave} we will adopt this SL method to study the s-wave insulator/superconductor phase transition, while p-wave case will be discussed in Sec.\ref{sec:pwave}. Finally, conclusions and discussions will be drawn in Sec.\ref{sec:conclusion}.

\section{Properties of the Sturm-Liouville Eigenvalue problems}
\label{sec:sl}
Consider a 2nd-order homogeneous ordinary differential equation (ODE) as,
\be \label{sleom}
\frac{d}{dx}\left(k(x)\frac{dy(x)}{dx}\right)-q(x)y(x)+\la \rho(x)y(x)=0,
\ee
in which, $x\in[a,b]$. For $x\in(a,b)$, $k(x)>0, q(x)\geq0, \rho(x)>0$. Multiply $y(x)$ to both sides of the above Eq.\eqref{sleom}, and make integration by parts, one can reach the following {\it Rayleigh Quotient} \cite{hilbert},
\be\label{quotient}
\la=\frac{-k(x)y(x)y'(x)\big|^b_a+\int^b_adx\left(k(x)y'(x)^2+q(x)y(x)^2\right)}{\int^b_a dx \rho(x)y(x)^2}
\ee
One of the assertions in the Sturm-Liouville eigenvalue problem states that if Eq.\eqref{sleom} satisfies the following boundary condition
\be\label{slbc}
k(x)y(x)y'(x)\big|^b_a=0,
\ee
 the ODE Eq.\eqref{sleom} will have infinite countable eigenvalues $\la_i$ with $0<\la_1<\la_2<\cdots$. This means the ODE will have a positive minimal eigenvalue and no maximal eigenvalues. The eigenfunctions $y_i$ corresponding to the eigenvalues $\la_i$ are complete and orthogonal. Actually, the eigenvalues $\la_i$ can be obtained from the extremal values of the following reduced {\it Rayleigh Quotient} Eq.\eqref{quotient},
\be\label{functional}
\mathfrak{K}[y(x)]=\frac{\int^b_adx\left(k(x)y'(x)^2+q(x)y(x)^2\right)}{\int^b_adx\rho(x)y(x)^2}.
\ee
In particular, the minimal eigenvalue $\la_1$ can be obtained from the minimum value of the above functional. In the following two sections, we will adopt the functional Eq.\eqref{functional} with the boundary condition Eq.\eqref{slbc} to estimate the critical chemical potentials for the most stable and the second stable modes of the holographic insulator/superconductor phase transition.

 \section{S-wave insulator/superconductor phase transition}
 \label{sec:swave}
The holographic s-wave insulator/superconductor phase transition was first studied in \cite{Nishioka:2009zj}, in which the model consists of an Einstein-Maxwell-scalar action with a negative cosmological constant $\Lambda$. The action is
\be
S=\int d^5x \sqrt{-G}\left(\frac{1}{16\pi G_N}(\mathcal{R}-\Lambda)+\frac{1}{g^2}\mathcal{L}\right),   \text{with}\quad \mathcal{L}=-\frac14F_{\mu\nu}F^{\mu\nu}-|D_\mu\psi|^2-m^2|\psi|^2.
\ee
where, $G_N$ is the gravitational Newton constant, $\mathcal{R}$ is the Ricci scalar, $\Lambda=-12/L^2$ while $L$ is the radius of the AdS spacetime. In addition, $F_{\mu\nu}=\partial_\mu A_\nu-\partial_\nu A_\mu$ is the U(1) field strength and $D_\mu=\partial_\mu-iqA_\mu$ is the covariant derivative. In the following we will work in the probe limit $G_N\to0$, which indicates  that the effect of Lagrangian of matter $\mathcal{L}$ to the gravitation can be neglected.

We will adopt the AdS soliton geometry\cite{Horowitz:1998ha} as the background, in which the line-element is
\be\label{adssoliton}
ds^2=L^2\frac{dr^2}{f(r)}+r^2(-dt^2+dx^2+dy^2)+f(r)d\chi^2.
\ee
where, $f(r)=r^2-r_0^4/r^2$. The background \eqref{adssoliton} has a compact spatial direction with $\chi\sim\chi+\pi L/r_0$ in order for the regularity of the geometry. For simplicity, the AdS soliton geometry can be obtained from double Wick rotation of the AdS Schwarzschild black hole\cite{Horowitz:1998ha,Wang:2012yj}, and it looks like a cigar where the tip is located at $r=r_0$. From \cite{Witten:1998zw} we know that AdS soliton is dual to a confined gauge field theory where a mass gap exists. Because of this mass gap which resembles the mass gap in the insulator in condensed matter physics, we can model the holographic insulator/superconductor phase transition in AdS soliton geometry. In particular, because there is no horizon in AdS soliton the temperature of it is zero.

For simplicity, we will only turn on the $t$-component of the gauge field and assume the scalar field be real. The ansatz is
\be A=\left(\phi(r), 0, 0, 0, 0\right),\quad \psi=\psi(r)=\psi^*(r).
\ee
Therefore, from the above action, line-element and the ansatz, we can readily get the equations of motions (EoMs) as:
\be\label{eompsi}
\frac{d^2\psi}{dr^2}+\left(\frac{\partial_rf}{f}+\frac3r\right)\frac{d\psi}{dr}+\left(-\frac{m^2}{f}+\frac{q^2\phi^2}{r^2f}\right)\psi&=&0,\\
\label{eomphi} \frac{d^2\phi}{dr^2}+\left(\frac{\partial_rf}{f}+\frac1r\right)\frac{d\phi}{dr}-\frac{2q^2\psi^2}{f}\phi&=&0.
\ee
The asymptotic behaviors of $\psi$ and $\phi$ near $r\to\infty$ are
\be\label{sexpansion}
\psi&\approx&\psi^{(1)}r^{-2+\sqrt{4+m^2}}+\psi^{(2)}r^{-2-\sqrt{4+m^2}}+\cdots,\nno\\
\phi&\approx&\mu-\rho r^{-2}+\cdots.
\ee
in which, from the dictionary of AdS/CFT correspondence, $\psi^{(i)} \quad (i=1, 2)$ are corresponding to the condensation values of the operator $O_{(i)}$ on the boundary field theory while $\mu$ and $\rho$ can be interpreted as chemical potential and charge density of the dual field theory, respectively. We will impose all the fields be finite at the tip $r=r_0$ rather than the ingoing boundary conditions near the horizon in black hole geometry.

There is a trivial solution $\psi=0$ to the above EoMs \eqref{eompsi} and \eqref{eomphi}, in this case the solution of $\phi$ will be
\be
\phi(r)=C_1+C_2 \frac{\tanh^{-1}({r^2}/{r_0^2})}{2r_0^2}.
\ee
In order for the $\phi(r)$ be regular at the tip $r_0$, one should impose $C_2\equiv0$ which mean $\phi\equiv\text{const}=\mu$ when $\psi=0$. However, there is another non-trivial solution when $\mu$ is greater than a critical value $\mu_c$. In this case, $\psi$ will develop a non-vanishing value in the bulk and it will spontaneously break the U(1) symmetry of the gauge field and then induce the superconducting phenomenon. This is the general concept of the holographic insulator/superconductor phase transition.

The critical point is just an onset of the phase transition, at which the scalar field $\psi$ is very small and the gauge field $\phi$ is nearly a constant. In the following, we are going to adopt the SL method \cite{Siopsis:2010uq,Cai:2011ky} to find the critical chemical potential $\mu_c$ of the phase transition analytically. Technically, we will set $\phi(r)$ be a constant $\mu$ rather than a dynamical field, meanwhile $\psi$ is still a dynamical field which satisfies Eq.\eqref{eompsi}. For convenience, in the rest parts of this section we will work in $z=1/r$ coordinate, then the EoM of $\psi$ Eq.\eqref{eompsi} becomes:
  \be
  \label{psiz}
  \psi ''(z)-\frac{\left(z^4+3\right) }{z-z^5}\psi
   '(z)+\frac{ \left(m^2-\mu ^2 z^2\right)}{z^2
   \left(z^4-1\right)}\psi (z)=0.
   \ee

 According to the dictionary of AdS/CFT correspondence, the conformal dimensions of the operator dual to the scalar field $\psi$ is $\Delta_{\pm}=2\pm\sqrt{4+m^2}$, in which $\Delta_{-}$ is corresponding to operator $O_{(1)}$ while $\Delta_{+}$ corresponding to $O_{(2)}$. It was pointed out \cite{Klebanov:1999tb} that when
 $0<\sqrt{4+m^2}<1\Rightarrow -4<m^2<-3$, there exist two different quantizations in the field theory which are related by a Legendre transformation. Explicitly, when $-4<m^2<-3$ the operator $O_{(1)}$ and $O_{(2)}$ are both normalizable, so we can regard each of $\psi^{(1)}$ or $\psi^{(2)}$ as the source and the other as the corresponding expectation value. However, when $m^2$ is out of this range only operator $O_{(2)}$ is normalizable while $O_{(1)}$ is non-normalizable. Therefore, we will investigate the critical behaviors of the phase transitions separately with respect to different condensations of the operator $O_{(i)}$ in the following two subsections.

\subsection{Condensation of operator $O_{(1)}$}
 \label{sect:O1}

 First, let us discuss the case in which operator $O_{(1)}$ condensates when $-4<m^2<-3$. Near the infinite boundary $r\to\infty\Rightarrow z\to0$, the expansion of $\psi$ can be read from Eq.\eqref{sexpansion} as, (We have set $\psi^{(2)}=0$, because it has the role as the source which we do not want to turn on.)
  \be\label{sexpansionf}
   \psi(z)|_{z\to0}\approx\psi^{(1)}z^{2-\sqrt{{m^2}+4}}\approx \langle O_1\rangle z^{2-\sqrt{{m^2}+4}}F(z),
  \ee
 where $\langle O_1\rangle$ is independent of $z$. We have introduced a trial function $F(z)$ into the expansion in order to estimate the critical value of the chemical potential of the phase transition.  It is obvious that at the infinite boundary $F(z)$ should satisfy $F(0)=1$.
  Therefore, substituting expansion \eqref{sexpansionf} into Eq.\eqref{psiz}, we reach,
  \be\label{FeomO1}
 && F''(z)+\frac{\left(\left(2 \sqrt{m^2+4}-5\right) z^4-2
   \sqrt{m^2+4}+1\right) }{z-z^5}F'(z)+\frac{
   \left(m^2-4 \sqrt{m^2+4}+8\right) z^2}{z^4-1}F(z)\nno\\&&-\frac{\mu
   ^2}{z^4-1}F(z)=0.\nno\\
  \ee
 Following the steps in \cite{Siopsis:2010uq,Cai:2011ky}, we can multiply
  \be
  T(z)=\left(1-z^4\right) z^{1-2 \sqrt{m^2+4}}
  \ee
  to both sides of Eq.\eqref{FeomO1} and write it in the standard SL method form as
   \be
\frac{d}{dz}\bigg(k(z)\frac{dF}{dz}\bigg)-p(z)F+q(z)\mu^2F=0.
\ee
where,
\be
k(z)=\left(1-z^4\right) z^{1-2 \sqrt{m^2+4}},\quad p(z)=\left(m^2-4 \sqrt{m^2+4}+8\right) z^{3-2 \sqrt{m^2+4}},\quad q(z)=z^{1-2 \sqrt{m^2+4}}.\nno\\
\ee
Please note that $k(z), p(z)$ and $q(z)$ satisfy the conditions of the SL method discussed in Sec.\ref{sec:sl}. Therefore, we can take advantage of the SL method to analytically estimate the eigenvalues of $\mu^2$. We notice that when $z\to0$, the leading order in $z$ for $k(z)$ is $s=1-2 \sqrt{m^2+4}$, in which $-1<s<1$ for the range of $-4<m^2<-3$. Therefore, in order to satisfy the condition Eq.\eqref{slbc}
\be\label{condition}
k(z)F(z)F'(z)\big|^1_0\approx 0-z^sF(0)F'(0)\big|_{z=0}=-z^sF'(0)\big|_{z=0}=0,
\ee
one should impose $F'(0)=0$. Therefore, we can assume $F(z)$ as the polynomial in $z$ as
\be F(z)=1-\al z^2,\ee
in which $\al$ is a constant parameter. It can be easily checked that the form of $F(z)$ above indeed satisfies the condition \eqref{condition} and the Dirichlet boundary condition $F(0)=1$. Therefore, the eigenvalues of $\mu^2$ can be achieved from the extremal values of the following functional by virtue of the {\it Rayleigh Quotient} \eqref{quotient},
\be
\mu^2&=&\frac{\int^1_0dz \left(kF'^2+pF^2\right)}{\int^1_0dz~qF^2}\nno\\
&=&\left(\sqrt{m^2+4}-3\right)
   \left(\sqrt{m^2+4}-2\right) \left(\sqrt{m^2+4}-1\right)
   \left(-40 \alpha ^2+2 \alpha  m^4-2 m^4-\alpha ^2
   \sqrt{m^2+4} m^2\right.\nno\\&&\left.+8 \alpha ^2 m^2-20 \alpha ^2
   \sqrt{m^2+4}+8 \alpha  \sqrt{m^2+4} m^2-16 \alpha
   m^2-5 \sqrt{m^2+4} m^2+10 m^2\right.\nno\\&&\left.+\alpha ^2 \sqrt{m^2+4}
   m^4-2 \alpha  \sqrt{m^2+4} m^4+\sqrt{m^2+4}
   m^4\right)\bigg/\bigg(m^2 \left(m^2-5\right) \left(6 \alpha ^2-14
   \alpha +\alpha ^2 m^2\right.\nno\\&&\left.-3 \alpha ^2 \sqrt{m^2+4}-2 \alpha
    m^2+8 \alpha  \sqrt{m^2+4}+m^2-5
   \sqrt{m^2+4}+10\right)\bigg).
\ee

\begin{figure}[h]
   \includegraphics[scale=.65]{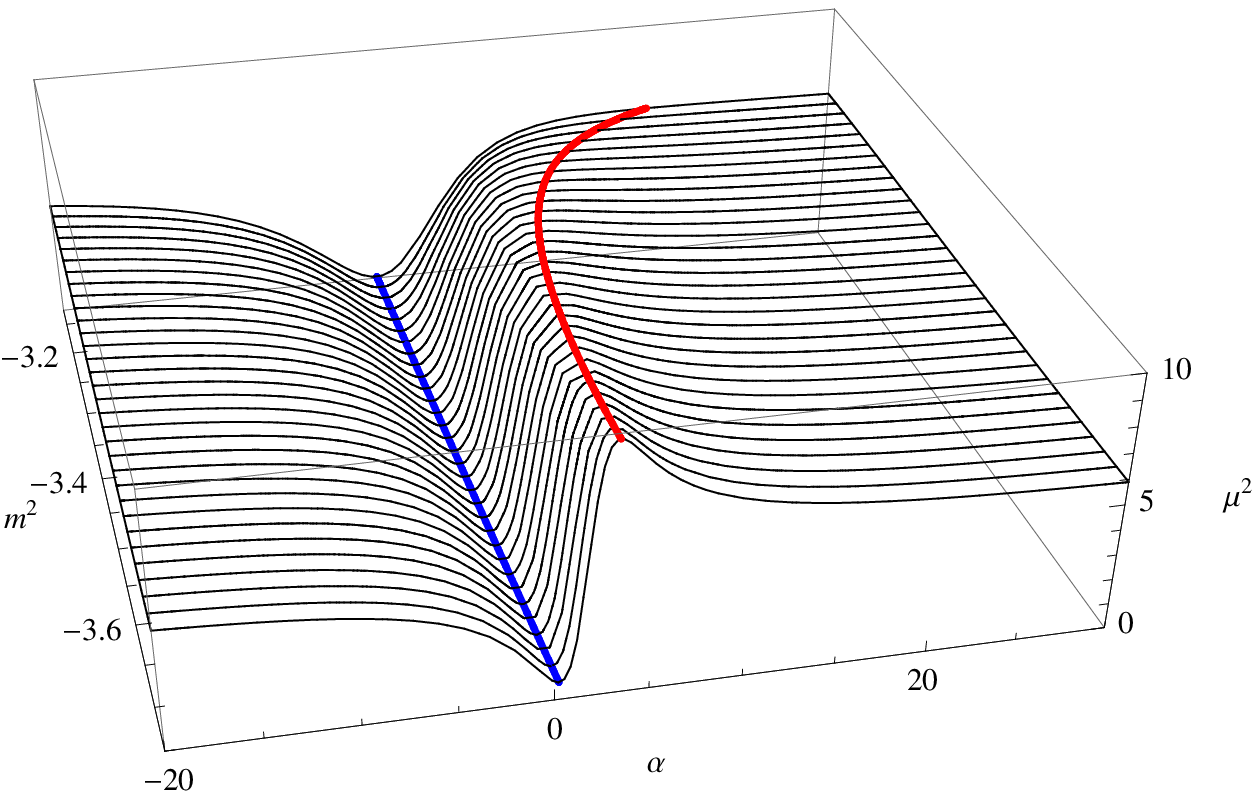}
    \includegraphics[scale=.55]{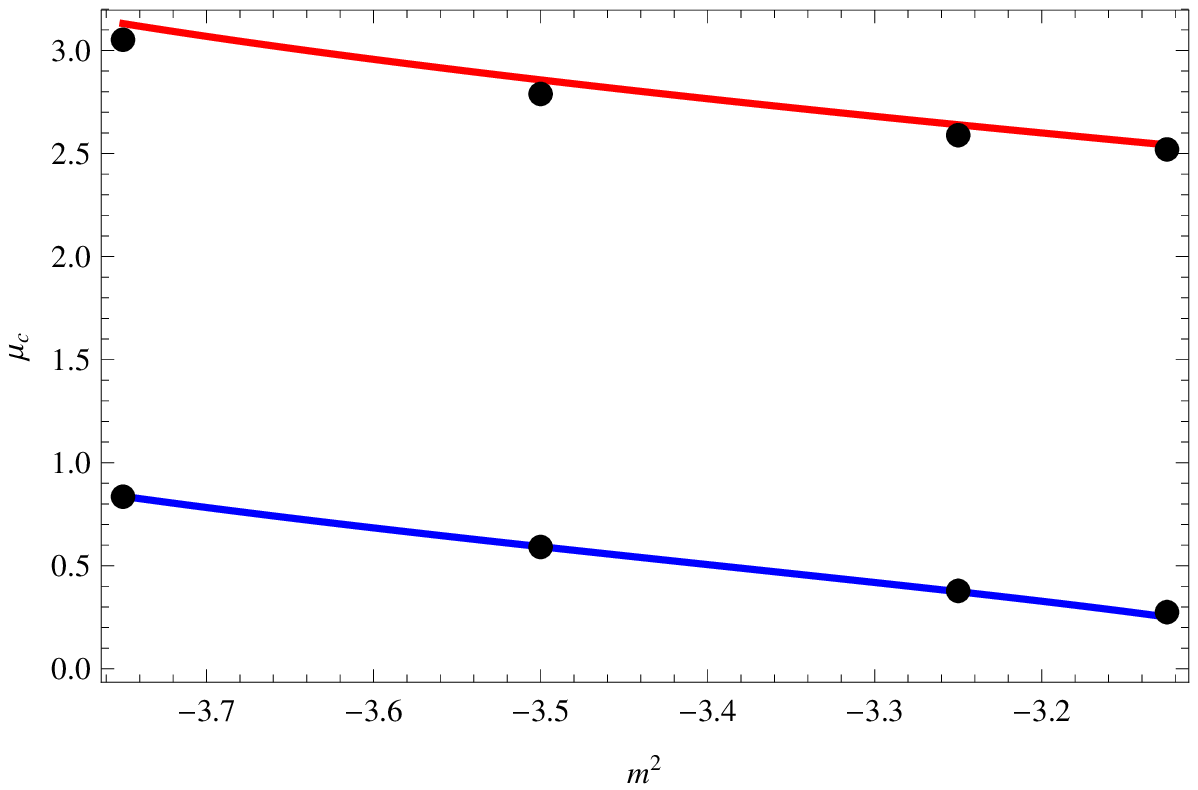}
  \caption{\label{3dO1} (Left)The 3D version of the chemical potential square $\mu^2$ as a function of $m^2$ and $\al$ for $\langle O_{(1)}\rangle$ condensation. The blue line shows the minimal values of $\mu^2$ while the red line illustrate the maximal values of it; (Right) Blue line and red line are extracted form the left panel, in which case $\mu_c$ is a one parameter function of $m^2$. Black dots are obtained from the shooting method.}
\end{figure}

 \begin{figure}[h]
   \includegraphics[scale=.68]{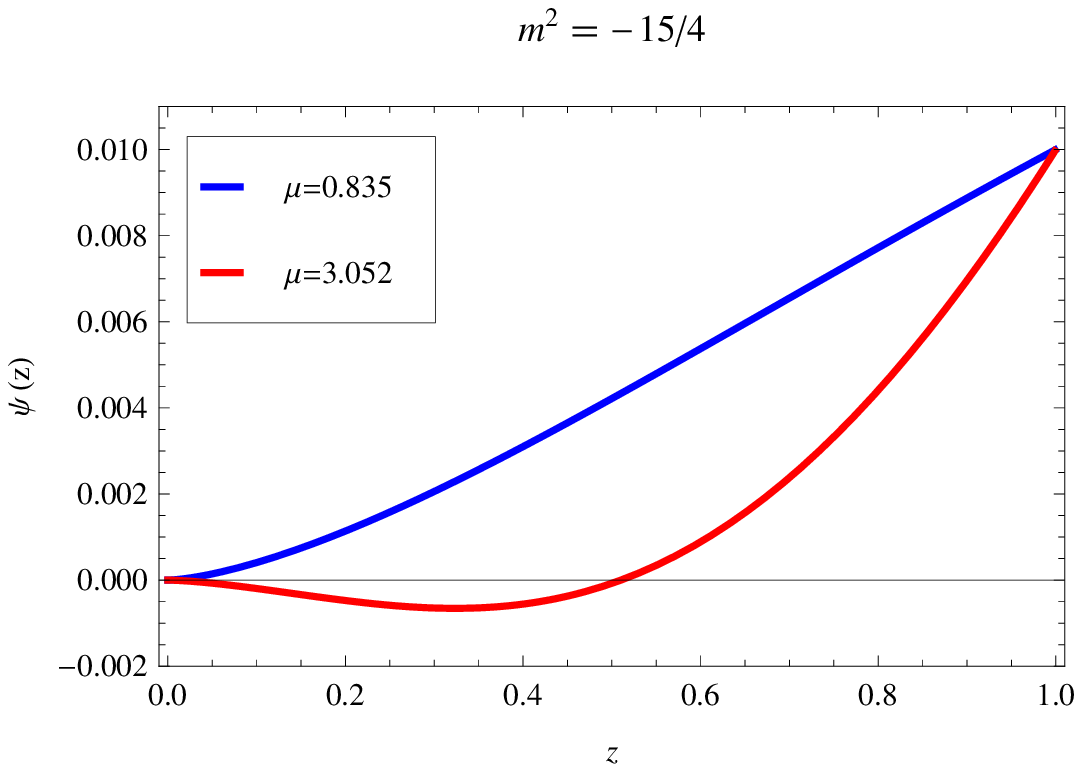}
   \hspace{-1.cm}\includegraphics[scale=.68]{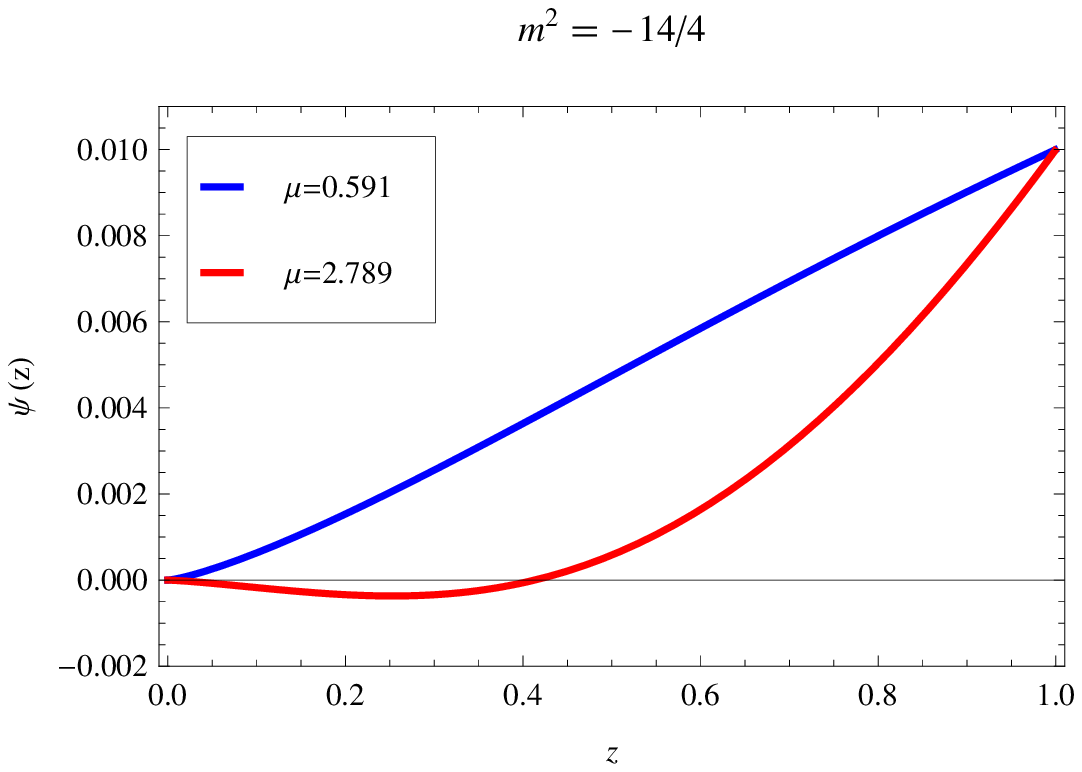}
   \includegraphics[scale=.68]{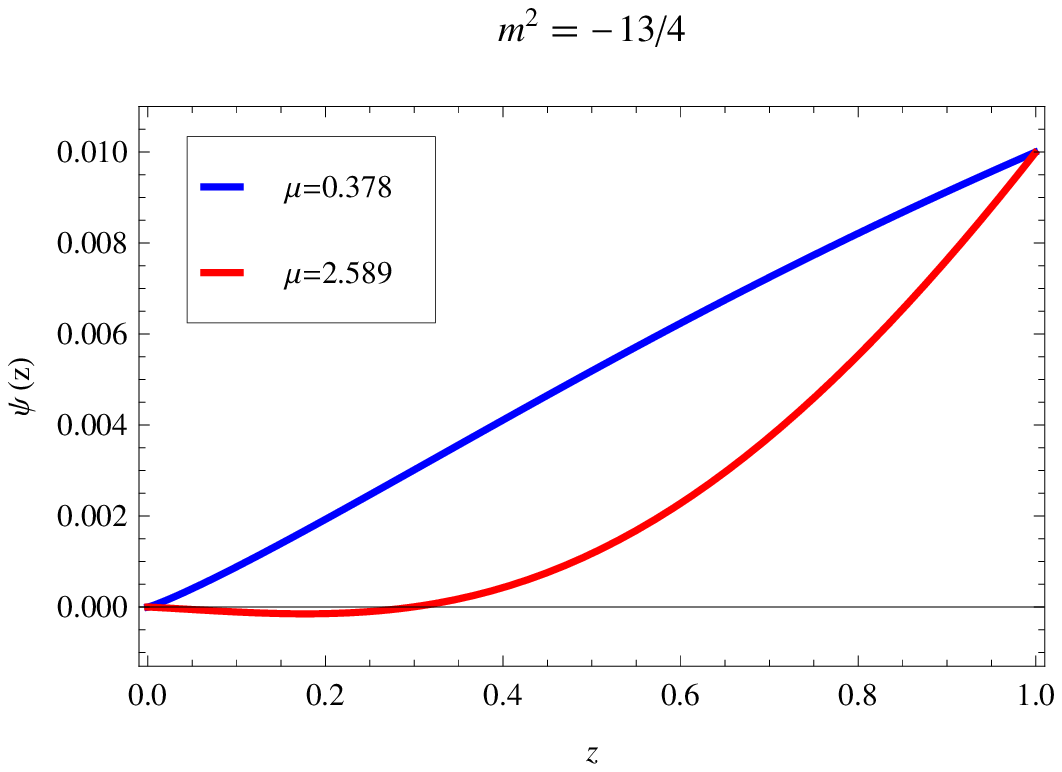}
   \hspace{-1.cm}\includegraphics[scale=.68]{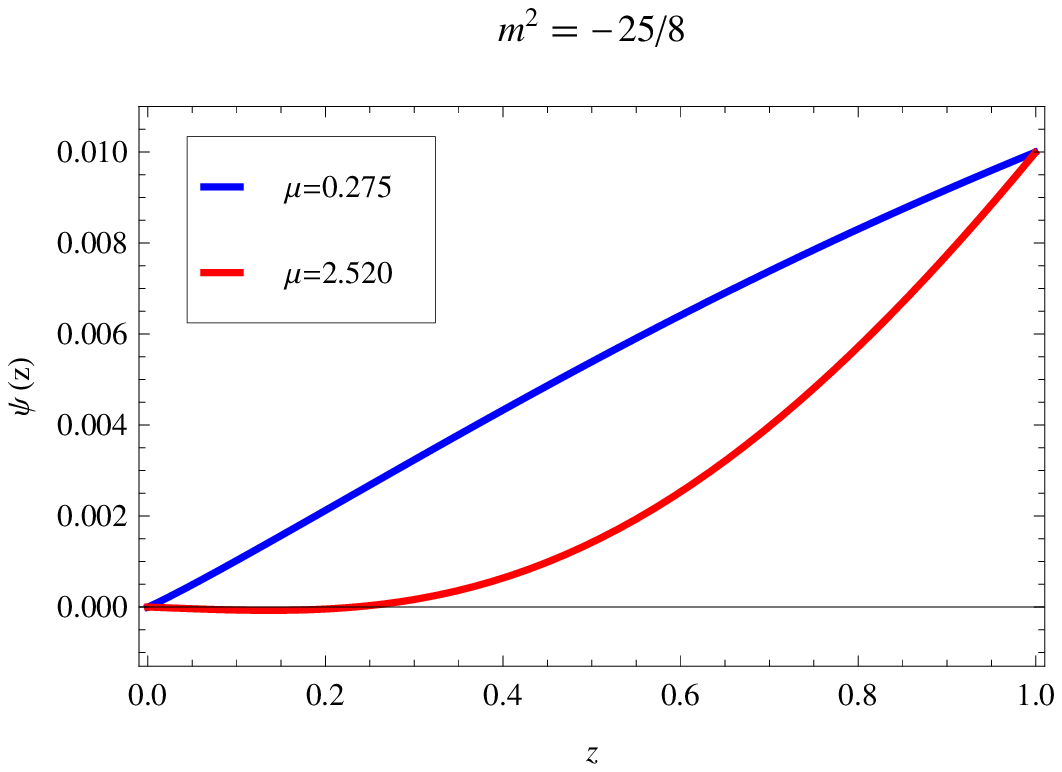}
  \caption{\label{O1psi} The profile of $\psi(z)$ with respect to various $m^2$ and critical chemical potential $\mu_c$ from shooting method for $\langle O_{(1)}\rangle$ condensation. The blue lines are corresponding to node zero solution while the red lines are corresponding to node one solution.}
\end{figure}

In the left panel of Fig.\ref{3dO1}, we plot the 3D version of $\mu^2$ as a function of $m^2$ and $\al$. The blue line is corresponding to the minimal values of $\mu^2$ while the red line is corresponding to the maximal values of $\mu^2$. Actually, these minimums and maximums of $\mu$ are corresponding to the critical chemical potentials of the most stable modes and second stable modes of the solutions, respectively \cite{Cai:2011qm}. Moreover, the critical chemical potentials can be written as one parameter functions of $m^2$, which are plotted in the right panel of Fig.\ref{3dO1}. In the right panel of Fig.\ref{3dO1}, the range of the $m^2$ is $-15/4\leq m^2\leq -25/8$ which lies in the range $-4<m^2<-3$. The black dots  are calculated from the shooting method to solve the EoMs \eqref{eompsi} and \eqref{eomphi} together by imposing $\psi^{(2)}=0$ at infinite boundary. The values of the black dots are corresponding to the values of the chemical potentials in Fig.\ref{O1psi}.

Therefore, the blue line here is related to the most stable mode of the solution, while the red line is related to the second stable mode. This is very similar to the marginally stable modes in \cite{Cai:2011qm}, in which the lowest lying mode is related to the blue line solution while the second lowest lying mode is related to the red line solution. The most stable mode is always regarded as node zero ($n=0$) solution, while the second stable mode is regarded as node one ($n=1$) solution from the language of QNMs. In Fig.\ref{O1psi}, we explicitly plot the profiles of $\psi$ with respect to various $m^2$ and $\mu_c$ from the numerical shooting method. We can intuitively see that $n=0$ solution has no intersecting points to the $\psi=0$ line, while $n=1$ solution has only one intersecting point to the $\psi=0$ line, except at $z=0$. The black dots in the right panel of Fig.\ref{3dO1} are exactly the values of the critical chemical potential $\mu_c$ obtained from Fig.\ref{O1psi}. We can find that $\mu_c$'s obtained from the SL method are perfectly in agreement with the ones got from the shooting method. The discrepancies of $\mu$'s obtained from these two approaches are within $2.57\%$. Therefore, we can see that the analytical SL method is not only suitable to estimate the critical chemical potential of the most stable mode, but also viable for the second stable mode.

As a bonus, we find that the critical chemical potential $\mu_c$ will decrease with respect to $m^2$, please refer to Fig.\ref{3dO1}. However, please keep in mind that the conformal dimension of operator $O_{(1)}$ is $\Delta_{-}=2-\sqrt{4+m^2}$. Therefore, actually $\mu_c$ will increase with respect to the conformal dimensions of the operator $O_{(1)}$.

 \subsection{Condensation of operator $O_{(2)}$}
 \label{sect:O2}

In this subsection, we will study the condensation for operator $O_{(2)}$ while setting $\psi^{(1)}=0$. The conformal dimension for $O_{(2)}$ is $\Delta_+=2+\sqrt{4+m^2}$. Following the steps in the previous subsection, we will introduce a trial function $F(z)$ into the expansion of $\psi$ as
\be
\psi|_{z\to0}\approx \langle O_{(2)}\rangle z^{2+\sqrt{4+m^2}}F(z),
\ee
It is apparent that $F(0)=1$. Thus the EoM of $\psi$ Eq.\eqref{psiz} becomes
\be
&&F''(z)+\frac{1+2\sqrt{4+m^2}-z^4(5+2\sqrt{4+m^2})}{z-z^5}F'(z)+\frac{(8+m^2+4\sqrt{4+m^2})z^2}{z^4-1}F(z)\nno\\&&+\frac{\mu^2}{1-z^4}F(z)=0.
\ee
Multiply $T(z)=(z^4-1)z^{1+2\sqrt{4+m^2}}$ to both sides of the above equation, we reach
\be
\frac{d}{dz}\bigg(k(z)\frac{dF}{dz}\bigg)-p(z)F+q(z)\mu^2F=0.
\ee
where
\be
k(z)=\left(1-z^4\right) z^{2 \sqrt{m^2+4}+1},\quad p(z)=\left(m^2+4 \sqrt{m^2+4}+8\right) z^{2 \sqrt{m^2+4}+3},\quad q(z)=z^{2 \sqrt{m^2+4}+1}.\nno\\
\ee
Please note that in this case, the leading order of $z$ in $k(z)$ near $z\to0$ is $s=1+2 \sqrt{m^2+4}$, in which $s\geq1$ for $m^2\geq-4$. Therefore, the boundary condition \eqref{slbc} is naturally satisfied from the properties of $k(z)$, {\it i.e.},
\be
k(z)F(z)F'(z)\big|^1_0=0,
\ee
Therefore, in this case we will just require $F(z)$ to satisfy the Dirichlet boundary condition $F(0)=1$ rather than imposing the Neumann boundary condition $F'(0)=0$ as in the above subsection or in \cite{Siopsis:2010uq,Cai:2011ky}. For simplicity, we can set
\be F(z)=1-\al z\ee
The eigenvalues of $\mu^2$ can be obtained from the extremal values of the following functional
\be
\mu^2&=&\frac{\int^1_0dz \left(kF'^2+pF^2\right)}{\int^1_0dz~qF^2}\nno\\
     &=&\left(\sqrt{m^2+4}+1\right) \left(\sqrt{m^2+4}+2\right) \left(2 \sqrt{m^2+4}+3\right) \left(900
   \alpha ^2-1920 \alpha +19 \alpha ^2 m^4-40 \alpha  m^4\right.\nno\\ &&\left.+21 m^4+87 \alpha ^2 \sqrt{m^2+4} m^2+286 \alpha
   ^2 m^2+450 \alpha ^2 \sqrt{m^2+4}-188 \alpha  \sqrt{m^2+4} m^2-616 \alpha  m^2\right.\nno\\&&\left.-960 \alpha
   \sqrt{m^2+4}+102 \sqrt{m^2+4} m^2+339 m^2+540 \sqrt{m^2+4}+2 \alpha ^2 \sqrt{m^2+4} m^4\right.\nno\\&&\left.-4 \alpha
   \sqrt{m^2+4} m^4+2 \sqrt{m^2+4} m^4+1080\right) \bigg/\bigg(\left(2 m^4+17 \sqrt{m^2+4} m^2+68 m^2\right.\nno\\&&\left.+135
   \sqrt{m^2+4}+270\right)\left(11 \alpha ^2-24 \alpha +2 \alpha ^2 m^2+5 \alpha ^2 \sqrt{m^2+4}-4 \alpha
   m^2-12 \alpha  \sqrt{m^2+4}\right.\nno\\&&\left.+2 m^2+7 \sqrt{m^2+4}+14\right)\bigg).
\ee

\begin{figure}[h]
   \includegraphics[scale=.50]{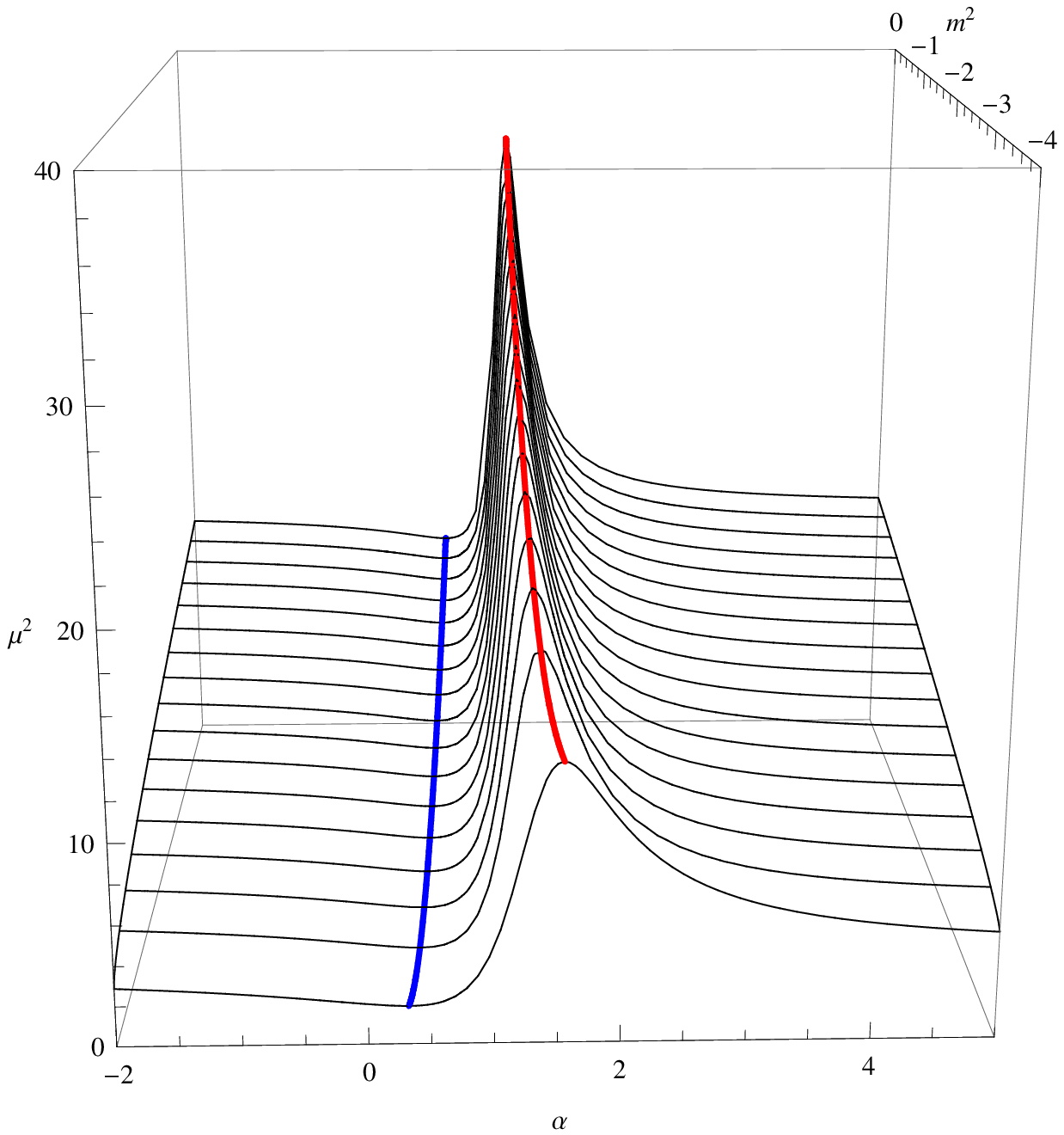}
   \includegraphics[scale=.65]{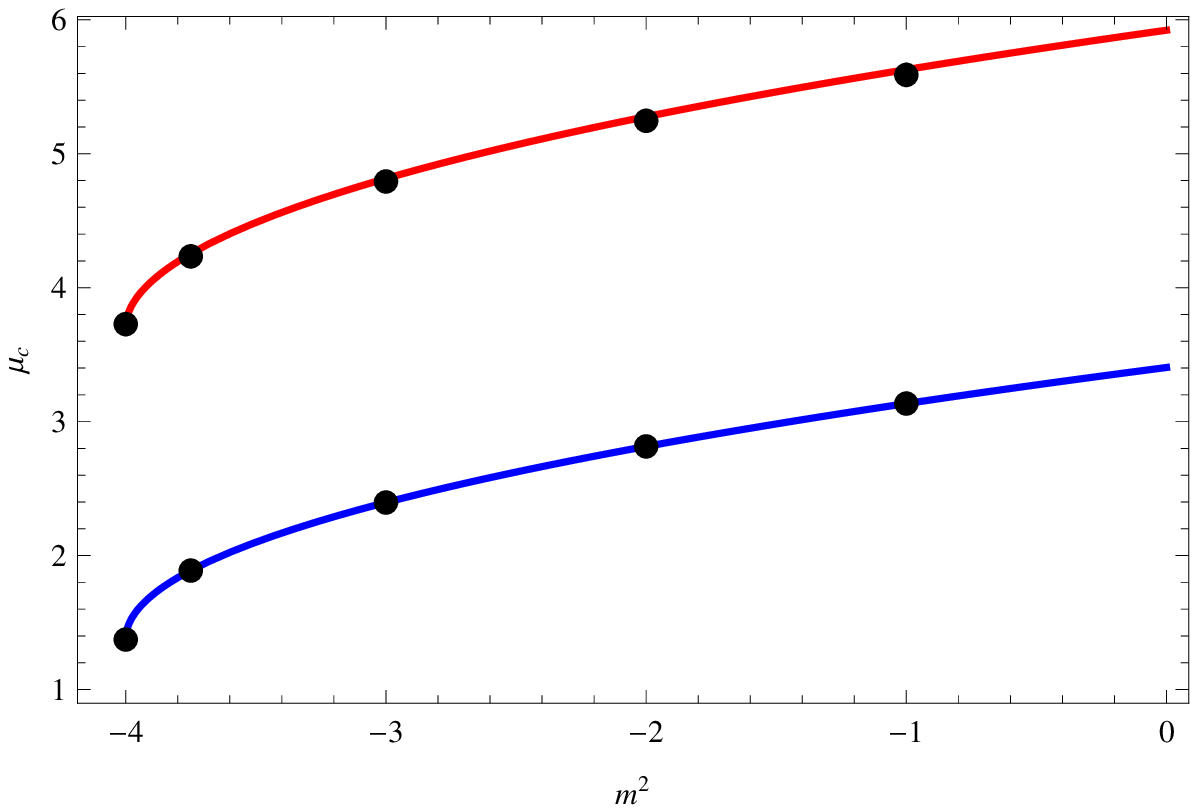}
  \caption{\label{3dO2}  (Left)The 3D version of the chemical potential square $\mu^2$ as a function of $m^2$ and $\al$ for $\langle O_{(2)}\rangle$ condensation. The blue line shows the minimal values of $\mu^2$ while the red line illustrate the maximal values of it; (Right) Blue line and red line are extracted form the left panel, in this case $\mu_c$ is a one parameter function of $m^2$. Black dots are obtained from the shooting method.}
\end{figure}

The profiles of functions $\mu^2(m^2,\al)$ are given in the left panel of Fig.\ref{3dO2}. The blue line is corresponding to the most stable mode of the solution while the red line is the second stable mode. On the right panel of Fig.\ref{3dO2} we show the critical chemical potential $\mu_c$ as a function of $m^2$, which is extracted from the left panel. The black dots are obtained from the shooting method in \cite{Cai:2011qm}. We can find that the values obtained from the two methods perfectly match each other with errors $0.70\%$. This tiny error again indicates that the SL method can not only find the most stable mode, but also can find the second stable mode just like the shooting methods did in \cite{Cai:2011qm}. In addition, this tiny error also suggests that the trial function $F(z)=1-\al z$ is very suitable to estimate the critical values of the phase transition. As a comparison, we also make use of the trial function $F(z)=1-\al z^2$ to estimate the second stable mode. We find that in this case the errors between it and the shooting method is up to $2.75\%$ which is bigger than $0.70\%$ obtained from $F(z)=1-\al z$. Therefore, we find that only requiring the  Dirichlet boundary condition of $F(z)$ is not only viable for estimating the critical chemical potentials, but also makes the estimations more precise. The reason for the more precision is that for the Dirichlet boundary condition $F(z)|_{z=0}=1$, the leading terms in the Taylor expansions of $F(z)$ is $1-\al z$ rather than $1-\al z^2$. In the next section, we give another exercise on p-wave insulator/superconductor phase transition in order to verify that the choice of $F(z)=1-\al z$ is better than $F(z)=1-\al z^2$ in estimating the critical values, if the boundary condition \eqref{slbc} is naturally satisfied by $k(z)$.

\section{p-wave insulator/superconductor phase transition}
\label{sec:pwave}
Following the setup in \cite{Cai:2011ky}, the p-wave insulator/superconductor phase transition can be studied by a SU(2) Einstein-Yang-Mills action with a negative cosmological constant. We will work in the probe limit, therefore we can only focus on the SU(2) Yang-Mills gauge field action, which is
\be
S_{YM}=\int d^5x \sqrt{-G}\left(-\frac14F^a_{\mu\nu}F^{a\mu\nu}\right).
\ee
in which $F^a_{\mu\nu}=\partial_\mu A^a_{\nu}-\partial_\nu A^a_\mu+\eps^{abc}A^b_\mu A^c_\nu$ is the Yang-Mills field strength, $a, b, c$ are the indices of the SU(2) Lie algebra generators, running from $1$ to $3$. $A=A^a_\mu \tau^a dx^\mu$, where $\tau^a$ are the generators of the SU(2) Lie algebra with commutations $[\tau^a, \tau^b]=\eps^{abc}\tau^c$. $\eps^{abc}$ is a totally antisymmetric tensor with $\eps^{123}=+1$.

We will adopt the conventions in \cite{Cai:2011ky} to denote the line-element of AdS soliton as (We have set $L=1, r_0=1$),
\be
ds^2=r^2\left(-dt^2+dx^2+dy^2+g(r)d\chi^2\right)+\frac{dr^2}{r^2g(r)}.
\ee
where $g(r)=1-1/r^4$. We can choose the gauge field as
\be
 A(r)=\phi(r)\tau^3dt+\psi(r)\tau^1dx.
\ee
In this ansatz, the gauge boson $\psi(r)$ is along $x$-direction which will be charged under $A^3_t=\phi(r)$. In this case, $\psi(r)$ is dual to a vector operator $O$ on the boundary field theory, while $\phi(r)$ is corresponding to the chemical potential. Therefore, the non-vanishing of $\psi$ will spontaneously break the U(1)$_3$ gauge symmetry, and then induce the superconducting phenomenon on the boundary.

From the above action, line-element and the ansatz of the fields, the EoMs of $\psi$ and $\phi$ are readily obtained as
\be\label{peompsi}
\frac{d^2\psi}{dr^2}+\left(\frac{\partial_rg}{g}+\frac{3}{r}\right)\frac{d\psi}{dr}+\frac{ \phi ^2}{r^4 g}\psi &=&0,\\
\label{peomphi}\frac{d^2\phi}{dr^2}+\left(\frac{\partial_rg}{g}+\frac{3}{r}\right)\frac{d\phi}{dr}-\frac{\psi^2 }{r^4 g}\phi &=&0.
\ee
The asymptotic expansions of $\psi$ and $\phi$ near $r\to\infty$ are
\be
\psi&\approx&\psi^{(0)}+\psi^{(2)}r^{-2}+\cdots,\\
\phi&\approx&\mu-\rho r^{-2}+\cdots.
\ee
where, from the dictionary of AdS/CFT correspondence, $\psi^{(0)}$ and $\psi^{(2)}$ are respectively the source and the expectation values of the dual operator $O$, while $\mu$ and $\rho$ can be interpreted of the chemical potential and charge density of the dual field theory, respectively. There is also a trivial solution of the above EoMs, which is $\psi=0$. In this case, the solution of $\phi(r)$ is
\be
\phi(r)=C_1-\frac{C_2}{4}\log\left(\frac{1-r^2}{1+r^2}\right).
\ee
In order for the regularity of $\phi$ at the tip $r=1$, one has to impose $C_2\equiv0$, {\it i.e.}, $\phi\equiv \text{const}=\mu$. Therefore, like what we did in the above section,  we will set $\phi=\mu$ and just consider the EoM of $\psi$ near the critical point of the phase transition. For convenience, we will work in $z=1/r$ coordinate. Thus the EoM of $\psi$ Eq.\eqref{peompsi} becomes
\be\label{ppsiz}
\psi ''(z)+\left(\frac{g'(z)}{g(z)}-\frac{1}{z}\right) \psi '(z)+\frac{\mu^2 \psi (z)}{g(z)}=0.
\ee
Again, introduce a trial function $F(z)$ into the expansions of $\psi$ near $z=0$ as
\be\label{pexp}
\psi|_{z\to0}\sim\langle O\rangle z^2F(z),
\ee
The boundary condition sets $F(0)=1$. Then substituting Eq.\eqref{pexp} into Eq.\eqref{ppsiz}, we reach
\be
F''(z)+\frac{\left(3-7 z^4\right) F'(z)}{z-z^5}+\frac{8 z^2 F(z)}{z^4-1}+\frac{F(z) \mu^2}{1-z^4}=0
\ee
Multiply $T(z)=z^3(1-z^4)$ to both sides of the above equation, the EoM becomes
  \be
\frac{d}{dz}\bigg(k(z)\frac{dF}{dz}\bigg)-p(z)F+q(z)\mu^2F=0.
\ee
where
\be
k=z^3(1-z^4),\quad p=8z^5,\quad q=z^3.
\ee
Considering that $k(z)$ will naturally render the boundary condition\eqref{slbc} satisfied, {\it i.e.},
\be
k(z)F(z)F'(z)\big|^1_0=0.
\ee
we need not impose any restrictions on $F'(z)$. Therefore, we simply assume that $F(z)=1-\al z$ to satisfy $F(0)=1$, then the eigenvalues of $\mu^2$ can be obtained from the extremal values of the following functional
\be
\mu^2=\frac{\int^1_0dz \left(kF'^2+pF^2\right)}{\int^1_0dz~qF^2}=\frac{5 \left(189 \alpha ^2-384 \alpha +224\right)}{14 \left(10 \alpha ^2-24 \alpha +15\right)}.
\ee

\begin{figure}[h]
   \includegraphics[scale=.8]{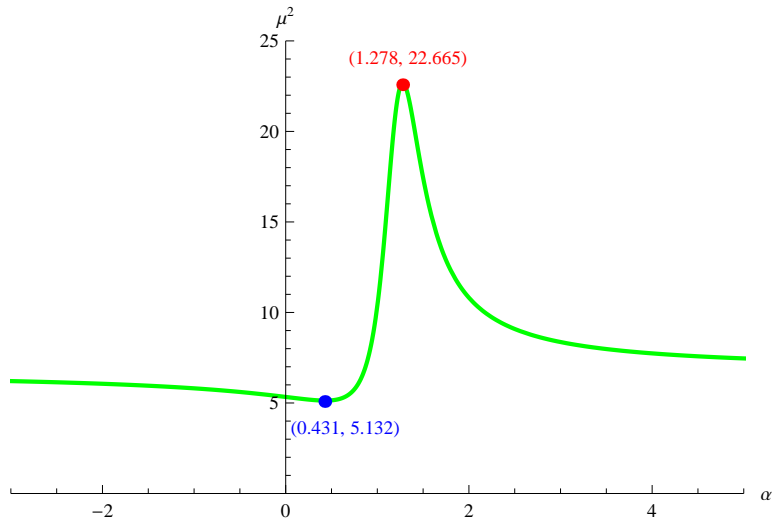}
  \caption{\label{mucpwave} The profile of $\mu^2$ as a function of $\al$. The blue point is corresponding to the most stable mode while the red point is corresponding to the second stable mode.}
\end{figure}

In Fig.\ref{mucpwave}, we plot the profile of $\mu^2$ as a function of $\al$. The blue point is corresponding the most stable mode to the solution, while the red point is corresponding to the second stable mode. It is easy to get that $\mu_{\text{min}}\sim\sqrt{5.132}\sim2.265$, when $\al=0.431$; $\mu_{\text{max}}\sim\sqrt{22.665}\sim4.760$, when $\al=1.278$. These two critical chemical potentials are in good agreement with the critical values in \cite{Cai:2011qm}, in which $\mu_c=2.265, 4.741$ for the first two lowest lying modes from the shooting method. We can see that the errors between the two methods are within $0.40\%$.  We also used the trial function $F(z)=1-\al z^2$ to estimate the critical chemical potential as a comparison, in which the errors are up to $2.59\%$ which is bigger than $0.40\%$. We again show that the SL method is suitable to estimate not only the most stable mode but also the second stable mode, moreover, the results obtained from the trial function $F(z)=1-\al z$ is more precise than those from $F(z)=1-\al z^2$ in previous literature \cite{Cai:2011ky}.

\section{Conclusions and Discussions}
\label{sec:conclusion}

In this paper, we took advantage of the analytical SL method to study the holographic insulator/superconductor phase transitions both for s-wave and p-wave, in the probe limit. We found that this analytical method could not only estimate the most stable (vacuum) mode, but also it could estimate the second stable mode. The results obtained from the SL method were perfectly in agreement with the numerical methods, such as the shooting method or the QNM method. \footnote{ {\blue Actually from \cite{Cai:2011qm} we know that the QNMs are actually the eigenvalues of the system. Besides, SL method is also an approach to calculate the eigenvalues. Therefore, theoretically there is a one-to-one corresponding relation between the eigenvalues from the SL method and from the QNM method, although in our draft we could just calculate up to the second lowest eigenvalue by introducing a trial function for estimation.}} Besides, we argued that actually one could relax the boundary conditions of the trial function $F(z)$, depending on whether the boundary condition $k(z)F(z)F'(z)\big|^1_0=0$ satisfied or not. If this boundary condition is automatically fulfilled by $k(z)$, we should only impose the Dirichlet boundary condition of $F(z)$ without the Neumann boundary conditions. We made use of two examples to verify that only imposing Dirichlet boundary condition of the trial function would be more precise than additionally imposing another Neumann boundary condition. We believe that this kind of SL method can be readily extended to study the holographic superconductor/metal phase transition in black hole geometry.

The problem of this analytical SL methods is that it could hardly find the third stable mode of the solution, like the shooting method and QNM method did in previous papers \cite{Cai:2011qm}. This is because the SL method states that the eigenvalues are countably infinite and have a sequence like $\la_1<\la_2<\cdots$. Therefore, the third stable mode should have a critical value $\mu_c^{(3)}$ which is greater than the first two modes $\mu_c^{(1)}$ and $\mu_c^{(2)}$. But from the profiles of $\mu$ or $\mu^2$, the denominator and the numerator of the {\it Rayleigh Quotient} are greater than zero,\footnote{Because in the {\it Rayleigh Quotient} \eqref{quotient}, $\rho(x)>0$ for $x\in(a,b)$, therefore, $\int^b_adx\rho(x)y(x)^2>0$ for $x\in[a,b]$. Therefore, the denominator of the {\it Rayleigh Quotient} is always positive. This argument goes the same way to the numerator of it.} therefore, $\mu$ is positive and analytical in $\al$ (please consult Fig.\ref{mucpwave}). So the next extremal value of $\mu$ should be smaller than $\mu_c^{(2)}=4.760$ (the red point) because the function of $\mu$ is a continuous and smooth function of $\al$. This contradicts with the assertion in SL method that $\la_3>\la_2$. Therefore, the analytical SL method for introducing a trial function depending on one parameter, such as $\al$, can only be applied to estimate the most and second stable mode of the solution. For higher modes, maybe one should introduce other types of the trial function depending on various parameters, or one should directly appeal to the shooting method or QNM method.

\section*{Acknowledgement}

H.F.L. would like to thank Prof.Rong-Gen Cai and Dr.Hai-Qing Zhang
for their indispensable discussions and comments. This work was
supported by the Young Scientists Fund of the National Natural
Science Foundation of China (Grant No.11205097), in part by the
National Natural Science Foundation of China (Grant Nos. 11075098,
11175109), Supported by Program for the Innovative Talents of Higher
Learning Institutions of Shanxi, and the Natural Science Foundation
for Young Scientists of Shanxi Province,China (Grant
No.2012021003-4) and by the Shanxi Datong University doctoral
Sustentation Fund(No. 2011-B-04), China.


\end{CJK*}
\end{document}